\documentclass[onecolumn]{emulateapj}
\usepackage{graphicx}
\begin{document}

\title{The Black Hole Mass Distribution in the Galaxy}

\author{Feryal \"Ozel$^1$, Dimitrios Psaltis$^1$, Ramesh Narayan$^2$, 
Jeffrey E. McClintock$^3$}

\affil{$^1$Department of Astronomy, University of Arizona, 933 N. 
Cherry Ave., Tucson, AZ 85721} 
\affil{$^2$ Institute for Theory and Computation, Harvard University, 
60 Garden St., Cambridge, MA 02138} 
\affil{$^3$ Harvard-Smithsonian Center for Astrophysics, 60 Garden St., 
Cambridge, MA 02138}

\begin{abstract}
We use dynamical mass measurements of 16 black holes in transient
low-mass X-ray binaries to infer the stellar black hole mass
distribution in the parent population. We find that the observations
are best described by a narrow mass distribution at $7.8 \pm
1.2~M_\odot$. We identify a selection effect related to the choice of
targets for optical follow-ups that results in a flux-limited
sample. We demonstrate, however, that this selection effect does not
introduce a bias in the observed distribution and cannot explain the
absence of black holes in the $2-5~M_\odot$~mass range. On the high
mass end, we argue that the rapid decline in the inferred distribution
may be the result of the particular evolutionary channel followed by
low-mass X-ray binaries. This is consistent with the presence of
high-mass black holes in the persistent, high-mass X-ray binary
sources. If the paucity of low-mass black holes is caused by a sudden
decrease of the supernova explosion energy with increasing progenitor
mass, this would have observable implications for ongoing transient
surveys that target core-collapse supernovae. Our results also have
significant implications for the calculation of event rates from the
coalescence of black hole binaries for gravitational wave detectors.
\end{abstract}


\section{Introduction}

The distribution of stellar black hole masses in the Galaxy is
intricately related to the population and evolution of massive stars,
the energetics and dynamics of supernova explosions, and the dividing
line between neutron stars and black holes. Inferring this
distribution from observations helps address a number of outstanding
questions in these areas.

Observations of mass losing evolved stars are one of the direct ways
of studying the end stages of stellar evolution (Smartt 2009).
Similarly, supernova and transient surveys (e.g., Rau et al. 2009 and
references therein) provide increasingly larger data samples that have
revealed a range of supernova properties and energetics that is wider
than previously anticipated. Studies of the end products of supernova
explosions, namely of the black hole and neutron star mass
distributions, provide a different and complementary approach to
constraining models of massive stellar populations, evolution, and
explosions (Timmes, Woosley, \& Weaver 1996; Fryer 1999; Fryer et al.\
2002; Woosley, Heger, \& Weaver 2002; Zhang, Woosley, \& Heger 2008).

The mass distribution of stellar black holes is a key ingredient in
the calculation of the event rates for gravitational wave
observatories such as LIGO, GEO500, and Virgo (Abadie et al. 2010). In
addition, inferring this distribution may shed light on the origins
and properties of pulsar-black hole binaries (Narayan, Piran, \& Shemi
1991), which are believed to be some of the best laboratories for
strong-field gravity tests (Wex \& Kopeikin 1999).

During the last two decades, the masses of a large number of black
holes in X-ray binaries have been measured (Remillard \& McClintock
2006; McClintock \& Remillard 2006), providing a sample suitable for
the statistical inference of their parent mass distribution. Using
this sample, we find in this article strong evidence for a very narrow
distribution of masses for black holes in transient low-mass X-ray
binaries. In particular, we show that there is a paucity of black
holes with masses $\simeq 2-5~M_\odot$. Our results confirm and
strengthen an earlier finding of Bailyn et al. (1998), who argued for
a low-mass gap based on a limited sample of black holes. We explore in
detail below known selection biases and demonstrate that the low-mass
gap cannot be attributed to the observational selection of targets. We
further argue that the lack of high-mass black holes in low-mass X-ray
binaries may result from a particular evolutionary path that leads to
their formation.

\section{Accreting Stellar-Mass Black Holes and Black Hole Candidates}

In Tables 1 and 2, we present basic data for all 23 confirmed
black-hole X-ray binaries (Remillard \& McClintock 2006).  For the
purpose at hand, we divide these systems into three groups: eight
transient systems with orbital periods exceeding one day; nine
shorter-period transient systems for which the measurement of black
hole mass is problematic; and six systems with persistent X-ray
sources and massive O/B-type secondaries.  In these tables, we provide
the following data for most of the systems: Galactic coordinates,
maximum X-ray intensity, orbital period, a distance estimate, mass
function, mass ratio, inclination angle, and black hole mass.  Our
primary focus is on the 17 transient systems listed in Tables~1 and 2.

We summarize in Table~3 data for thirty-two additional transient
systems that are believed to contain black hole primaries based on the
spectral and timing properties of their X-ray sources.  For each of
these systems, we give both celestial and Galactic coordinates and the
maximum X-ray intensity that has been reported.

\subsection{The Twenty-Three Black Hole Binaries}

We now survey the constraints that have been placed on the masses of
the 23 confirmed stellar black holes via dynamical measurements. In
determining the masses of black holes in X-ray binaries, the mass
function 
\begin{equation}
f(M) \equiv \frac{P_{\rm orb}K^{3}}{2\pi G}=\frac{M \sin^3 i}{(1+q)^{2}}
\end{equation}
is the most important and secure observable. The orbital period
$P_{\rm orb}$ and the half-amplitude of the velocity curve of the
secondary star $K$ can, in most cases, be determined precisely and
accurately (Remillard \& McClintock 2006; Charles \& Coe 2006).  These
two quantities define the value of the mass function, which is an
absolute lower limit on the mass of the compact object: i.e., $M \ge
f(M)$. The mass function relates the black hole mass $M$, the orbital
inclination angle $i$, and the mass ratio $q~\equiv M_{2}/M$, where
$M_{2}$ is the mass of the secondary star.  Values of the mass
function, the inclination, and the mass ratio for the 23 established
black hole binaries are given in Table 2. The mass-ratio estimates
were obtained or derived from Orosz (2003) and Narayan \& McClintock
(2005), with a few refinements based on the references cited in the
table. Determining the black hole mass $M$, the quantity of interest,
is challenging because in many cases it is difficult to obtain secure
constraints on the inclination and the mass ratio.

Table 1 separates the 23 systems into three groups based on their
X-ray behavior: the persistent sources (PS), the long-period
transients (LPT), and the short-period transients (SPT).  This is
important for understanding possible selection effects, which we
explore in Section~4. Table~2, on the other hand, separates sources
into three different groups based on the amount of data available on
their mass ratios and inclinations. This impacts the inference of the
masses of individual sources, which we discuss in Section~3. These
tables sparely give for each source a few key references, which are
supplemented by some additional references in the discussion that
follows.

\subsubsection{The Persistent Sources}

These systems contain O/B-type secondaries and are persistently X-ray
bright.  Five of them have relatively massive black-hole primaries,
$M>10~M_{\odot}$.  However, the mass of LMC X-3 is presently poorly
constrained: $4~M_{\odot}~\le~M~\le~11M_{\odot}$ (Cowley 1992).  In
the case of Cyg X-1, a very wide range of masses down to $5~M_{\odot}$
has recently been considered (Caballero-Nieves et al.\ 2009).
However, recent VLBA observations have shown that the distance exceeds
1.5~kpc (M.\ Reid, private communication), which firmly establishes
$M>8~M_{\odot}$.  There are two caveats on the mass constraints on IC
10 X-1 and NGC 300-1: These results are based on less-reliable
emission-line radial velocities and assume specific lower bounds on
the masses of the secondary stars.

\subsubsection{The Long Period Transient Sources}

Nearly all transient black hole sources, which are fed by Roche-lobe
overflow, exhibit long periods of deep quiescence during which the
spectra of their secondary stars are prominent.  For all eight of
these long-period systems ($P_{\rm orb}>1$~d) there is strong evidence
that the masses of their black hole primaries exceed $6~M_{\odot}$.
We comment on the two weakest cases.  (1) GX 339-4 has never reached a
deep enough quiescent state to reveal its photospheric absorption
lines, and its mass function was determined via the Bowen emission
lines (Hynes et al.\ 2003).  Furthermore, this system does not exhibit
ellipsoidal light curves that allow its inclination to be constrained.
For a defense of the $M=6~M_{\odot}$ lower limit see Munoz-Darias et
al.\ (2008).  (2) GRS 1915+105: The pioneering mass measurement of
this system by Greiner et al.\ (2001) requires confirmation.  Their
spectroscopic orbital period disagrees by 8\% with a more recent
photometric determination of the period (Neil et al.\ 2007).  Again,
there are no well-behaved ellipsoidal light curves that can be used to
constrain the inclination, which in this case is inferred from a
kinematic model of the relativistic radio jets (Mirabel \&
Rodr{\'{\i}}guez 1999).

\subsubsection{The Short Period Transient Sources}

Significant constraints have been placed on the black hole primaries
of only three of the nine short period systems ($P_{\rm orb}<1$~d).
It is difficult to obtain reliable inclination constraints for these
systems because studies in quiescence of their small, late-type
secondaries are compromised by the presence of a relatively strong and
variable component of non-stellar light (e.g., Zurita et al.\ 2003;
Cantrell et al.\ 2008), which is continually present in all of these
systems.  

A measurement of the inclination angle has been obtained only for
A0620--00, the prototype system, which is two magnitudes brighter than
the other short period transients.  The study by Cantrell et al.\
(2010), a tour de force that makes use of 32 photometric data sets
spanning 30 years, relies on 10 data sets obtained when the source was
in a ``passive'' quiescent state (Cantrell et al.\ 2008).  As
indicated in Table 2, the authors constrain the mass to be
$M=6.6\pm0.25~M_{\odot}$.  Meanwhile, the masses of three other
systems (XTE J1118+480, Nova Oph 1977 and GS 2000+251) have by dint of
their large mass functions been shown to exceed $6~M_{\odot}$ (Table
2).  The large mass function of XTE J1859+226 (Table 2) suggests that
it also should be included in this group; however, this result is
unreliable, having only been presented in an IAU Circular (Filippenko
\& Silverman 2001), and the orbital period is uncertain (Zurita et
al.\ 2002). We exclude this source from further considerations on
deriving the black-hole mass distribution.

Among the 23 systems in Table 2, arguably the best candidate for
hosting a low-mass black hole is GRO J0422+32.  However, by briefly
discussing recent attempts to determine the system's inclination and
mass, we show that current results are presently unreliable: Beekman
et al.\ (1997) constrained the inclination to lie in the range
$i=10^{\circ}-31^{\circ}$ and concluded $M>9~M_{\odot}$.  Webb et al.\
(2000) concluded $i<45^{\circ}$ and $M>2.2~M_{\odot}$ with a maximum
mass that is not constrained.  Gelino and Harrison (2003) present the
strongest evidence for a low mass; they conclude $i>43^{\circ}$ and
$M<4.92~M_{\odot}$.  However, in a more recent Keck $K$-band study
Reynolds et al.\ (2007) found that ``No clear ellipsoidal modulation
is present in the light curve...'' and concluded, ``...that previous
infrared-based attempts to constrain the mass of the putative black
hole in this system are prone to considerable uncertainty.''  Thus, it
appears that a far more comprehensive photometric study (cf. Cantrell
et al.\ 2010) is required in order to obtain a firm mass constraint.

There are six systems in Table 2 with black holes of indeterminate
mass. The prospects for measuring or usefully constraining the masses
of two of them, LMC X-3 and XTE J1859-226, are bright: Work on the
former is almost complete (J.\ Orosz, private communication), and the
additional data required to confirm the orbital period and large mass
function of the latter are obtainable.  That leaves four systems (GRO
J0422+32, GRS 1009--45, Nova Mus 1991 and XTE J1650--500) with black
hole masses that are only very weakly constrained: $M>~f(M)~\gtrsim
1-3~M_{\odot}$.  As indicated above and as Cantrell et al.\ (2010)
have shown, it will be very challenging to place stronger and reliable
constraints on the masses of these black holes.

\subsection{Thirty-two Transient Black-Hole Candidates}

Data for 32 X-ray transient systems are given in Table~3.  These
systems lack radial velocity data, and most even lack an optical
counterpart.  Thus, presently, there are no dynamical constraints on
the masses of their compact primaries, which are believed to be black
holes because they share certain characteristic X-ray properties with
the 23 established black holes (McClintock \& Remillard 2006).  As
indicated in Table 3, the primary source of information about these
systems is the catalogue of Liu et al.\ (2007) and references therein.
For additional information and references on many of these systems, see
Table 4.3 and text in McClintock and Remillard (2006).

%%%%%%%%%%%%%%%%%%%%%%
%%%%   Table 1  %%%%%%
%%%%%%%%%%%%%%%%%%%%%%

\begin{deluxetable}{llllllllll}
\tabletypesize{\scriptsize}
%\rotate
\tablewidth{0pt}
\tablenum{1}
\tablecaption{Properties of Twenty-Three Black Hole Binaries}
\tablehead{
 \colhead{} &
 \colhead{Coordinate} &
 \colhead{Common Name} &
 \colhead{Type\tablenotemark{b}} &
 \colhead{$l$} &
 \colhead{$b$} &
 \colhead{Max.\ Int.} &
 \colhead{$P_{\rm orb}$} &
 \colhead{$D$} &
 \colhead{References\tablenotemark{c}} 
\cr
 \colhead{} &
 \colhead{Name} &
 \colhead{or Prefix\tablenotemark{a}} &
 \colhead{} &
 \colhead{(deg)} &
 \colhead{(deg)} &
 \colhead{(Crab)} &
 \colhead{(hr)} &
 \colhead{(kpc)} &
 \colhead{(distance)}
}
\startdata
% Table: Data for 23 Black Hole Binaries
%
%   Coordinate    Common Name  Type   l        b           Max. Int. P_orb   Distance       Ref.
%   Name          or Prefix           (deg)    (deg)      (Crab)     (hr)     (kpc)         (Distance)
%------------------------------------------------------------------------------------------------------------------------------------------------------------------------

1&  1354-64&      (GS)&        LPT&   310.0&   -2.8&       0.12&     61.1&   $>25$&           1\\
2&  1543-47&      (4U)&        LPT&   330.9&   +5.4&       15&       26.8&   $7.5\pm0.5$&     2\\
3&  1550-564&     (XTEJ)&      LPT&   325.9&   -1.8&       7.0&      37.0&   $4.4\pm0.5$&     3\\
4&  1655-40&      (GROJ)&      LPT&   345.0&   +2.5&       3.9&      62.9&   $3.2\pm0.5$&     4\\
5&  1659-487&     GX 339-4&    LPT&   338.9&   -4.3&       1.1&      42.1&   $9\pm3$&         5\\
6&  1819.3-2525&  V4641 Sgr&   LPT&   6.8&     -4.8&       13&       67.6&   $9.9\pm2.4$&     6\\
7&  1915+105&     (GRS)&       LPT&   45.4&    -0.2&      3.7&      739&    $9\pm3$&         7\\
8&  2023+338&     (GS)&        LPT&   73.1&    -2.1&       20&       155.3&  $2.39\pm0.14$&   8\\

\hline

9&  0422+32&      (GROJ)&      SPT&   166.0&   -12.0&      3&        5.1&    $2\pm1$&         9,10\\
10& 0620-003&     (A)&         SPT&   210.0&   -6.5&       50&       7.8&    $1.06\pm0.12$&   11\\
11& 1009-45&      (GRS)&       SPT&   275.9&   +9.4&       0.8&      6.8&    $3.82\pm0.27$&   10\\
12& 1118+480&     (XTEJ)&      SPT&   157.6&   +62.3&      0.04&     4.1&    $1.7\pm0.1$&     12\\
13& 1124-683&     Nova Mus 91& SPT&   295.3&   -7.1&       3&        10.4&   $5.89\pm0.26$&   10\\
14& 1650-500&     (XTEJ)&      SPT&   336.7&   -3.4&       0.6&      7.7&    $2.6\pm0.7$&     13\\
15& 1705-250&     Nova Oph 77& SPT&   358.2&   +9.1&       3.6&      12.5&   $8.6\pm2.1$&     14\\   
16& 1859+226&     (XTEJ)&      SPT&   54.1&    +8.6&       1.5&      9.2\tablenotemark{d}&   $8\pm3$&  10\\
17& 2000+251&     (GS)&        SPT&   63.4&    -3.0&       11&       8.3&    $2.7\pm0.7$&     14\\

\hline

18& 0020+593&     IC 10 X-1&   PS&    \nodata& \nodata&    0.00006&  34.9&   \nodata&        \nodata \\
19& 0055-377&     NGC 300-1&   PS&    \nodata& \nodata&    0.00004&  32.3&   \nodata&        \nodata \\
20& 0133+305&     M33 X-7&     PS&    \nodata& \nodata&    0.00002&  82.9&   \nodata&        \nodata \\
21& 0538-641&     LMC X-3&     PS&    \nodata& \nodata&    0.06&     40.9&   \nodata&        \nodata \\             
22& 0540-697&     LMC X-1&     PS&    \nodata& \nodata&    0.03&     93.8&   \nodata&        \nodata \\
23& 1956+350&     Cyg X-1&     PS&    \nodata& \nodata&    2.3&      134.4&  \nodata&        \nodata       

\enddata
\scriptsize

\tablenotetext{a} {The entries in parentheses are prefixes to the
  coordinate names that identify the discovery X-ray mission.}
\tablenotetext{b} {PS = persistent source; LPT = long period transient; SPT =
  short period transient.}
\tablenotetext{c} {References:
1.\ Casares et al.\ 2009;
2.\ J.\ Orosz, private communication;
3.\ Orosz et al.\ 2010;
4.\ Hjellming \& Rupen 1995;
5.\ Hynes et al.\ 2004;
6.\ Orosz et al.\ 2001;
7.\ Fender et al.\ 1999;
8.\ Miller-Jones et al.\ 2009;
9.\ Webb et al.\ 2000;
10.\ Hynes et al.\ 2005;
11.\ Cantrell et al.\ 2010;
12.\ Gelino et al.\ 2006;
13.\ Homan et al.\ 2006;
14.\ Barret et al.\ 1996}
\tablenotetext{d} {Unconfirmed and uncertain value; see text.}
\mbox{}
\end{deluxetable}

%%%%%%%%%%%%%%%%%%%%%%%%%%%%%%%%%%%%%%%%%%%%%%%%%%%%%%%%%%%%%%%%%%%%%%

\section{Black Hole Mass Measurements and Constraints}

In this section, we use the measurements of the mass functions, as
well as any available constraints on the mass ratios and inclinations
for the black holes in low-mass X-ray binaries shown in Table~2 in
order to place quantitative constraints on the individual black hole
masses. In particular, our aim is to derive the likelihood $P_i({\rm
data} \vert M)$, which measure the chance of obtaining the particular
set of data shown in Table~2 for the $i$-th source if that source had
mass $M$.

We divide the sources into three categories based on the amount and
quality of information regarding their mass ratios and inclinations:

\noindent {\em (i)\/} For six sources, the mass ratios and the inclinations
are tightly constrained, leading to well-determined black hole masses.
In this case, the probability distribution can be described as a
Gaussian
\begin{equation}
P_i({\rm data} \vert M) = C_i \exp\left[\frac{-(M-M_{0,i})^2}
{2\sigma_{M,i}^2} \right]
\end{equation}
with a mean $M_{0,i}$ and a standard deviation $\sigma_{M,i}$. In this
and the following expressions, $C_i$ is a proper normalization
constant such that
\begin{equation}
\int_0^\infty P_i({\rm data} \vert M) dM = 1.
\end{equation} 
This category includes A0620$-$003, 4U~1543$-$47, XTE~J1550$-$564,
GRO~J1655$-$40, V4641 Sgr, and GS~2023+338. 

\noindent {\em (ii)\/} For the sources in the second category, there is 
only a measurement of the mass function and constraints on the mass
ratio $q$. Here, we assume a Gaussian probability distribution over
the mass function with a mean $f_{0,i}$ and a standard deviation
$\sigma_{0,i}$. For the mass ratio, we adopt a uniform distribution
\begin{equation}
P(q) dq = \frac{dq}{q_{\rm max}-q_{\rm min}} 
\end{equation}
between the minimum and maximum allowed mass ratios, $q_{\rm min}$ and
$q_{\rm max}$, respectively. For each value of the mass ratio, the lack of 
eclipses implies a maximum value of the inclination, i.e., a minimum value 
of $\cos i$, such that
\begin{equation}
(\cos i)_{\rm min} = 0.462 \left(\frac{q}{1+q}\right)^{1/3}.
\end{equation}
Assuming a uniform distribution over $\cos i$ subject to this constraint, 
i.e., 
\begin{equation}
P(\cos i \vert q) d(\cos i) = \frac{d(\cos i)}{1-(\cos i)_{\rm min}}, 
\quad (\cos i)_{\rm min} \leq \cos i \leq 1, 
\end{equation}
yields
\begin{equation}
P_i({\rm data} \vert M) = C_i \int_{q_{\rm min}}^{q_{\rm max}} dq 
\int_{(\cos i)_{\rm min}}^1 \frac{d(\cos i)}{1-(\cos i)_{\rm min}}
\exp \left\{- \frac{[f_{0,i}-M \sin^3 i/(1+q)^2]^2} {2 \sigma_{f,i}^2}
\right\}.
\label{eq:prob2}
\end{equation}

The following nine sources belong to this category: GROJ 0422+32,
GRS~1009-45, XTE~J1118+480, Nova Mus 91, MS~1354-64, XTE~J1650-500,
GX~339-4, Nova Oph 77, and GS~2000+251.

\noindent {\em (iii)\/} This last category includes only GRS~1915+105, 
for which the mass function and the inclination have been measured,
and the mass ratio has been constrained (see the discussion in
Section~2 about the inclination measurement). In this case, we
calculate the probability distribution over mass using
equation~(\ref{eq:prob2}), supplemented by a Gaussian distribution
over inclination
\begin{equation}
P_i({\rm data} \vert M) = C_i \int_{q_{\rm min}}^{q_{\rm max}} dq 
\int_{(\cos i)_{\rm min}}^1 \frac{d(\cos i)}{1-(\cos i)_{\rm min}}
\exp \left\{- \frac{[f_{0,i}-M \sin^3 i/(1+q)^2]^2} {2 \sigma_{f,i}^2}
- \frac{(i-i_0)^2}{2 \sigma_i^2}
\right\}.
\label{eq:prob3}
\end{equation}

\begin{figure}
\centering
   \includegraphics[scale=0.75]{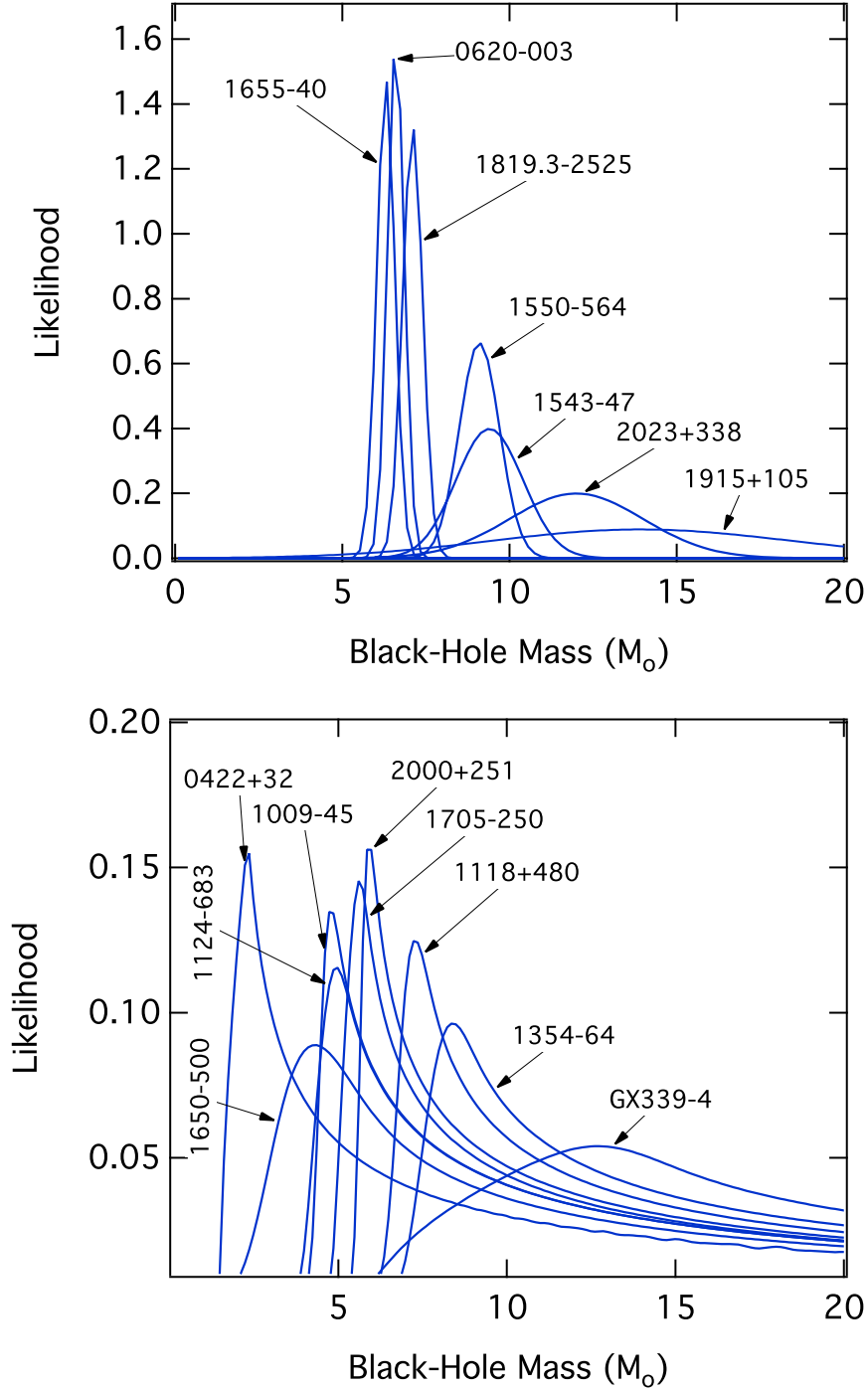}
\caption{The likelihoods $P_i({\rm data}\vert M)$ for the 16
sources in low-mass X-ray binaries that have been securely identified
as black holes. The top panel includes sources in categories {\em (i)}
and {\em (iii)}, while the bottom panel shows those in category {\em
(ii)}. The categories are based on the amount of information available
on the mass ratios and inclinations of the black hole binaries and are
discussed in more detail in the text.}
\label{fig:bh_prob} 
\end{figure}

Figure~\ref{fig:bh_prob} shows the likelihoods $P_i({\rm data}\vert M)$
for the 16 sources in the above three categories. The top panel
includes sources in categories {\em (i)} and {\em (iii)}, while the
bottom panel shows those in category {\em (ii)}.

\begin{figure}
\centering
   \includegraphics[scale=0.75]{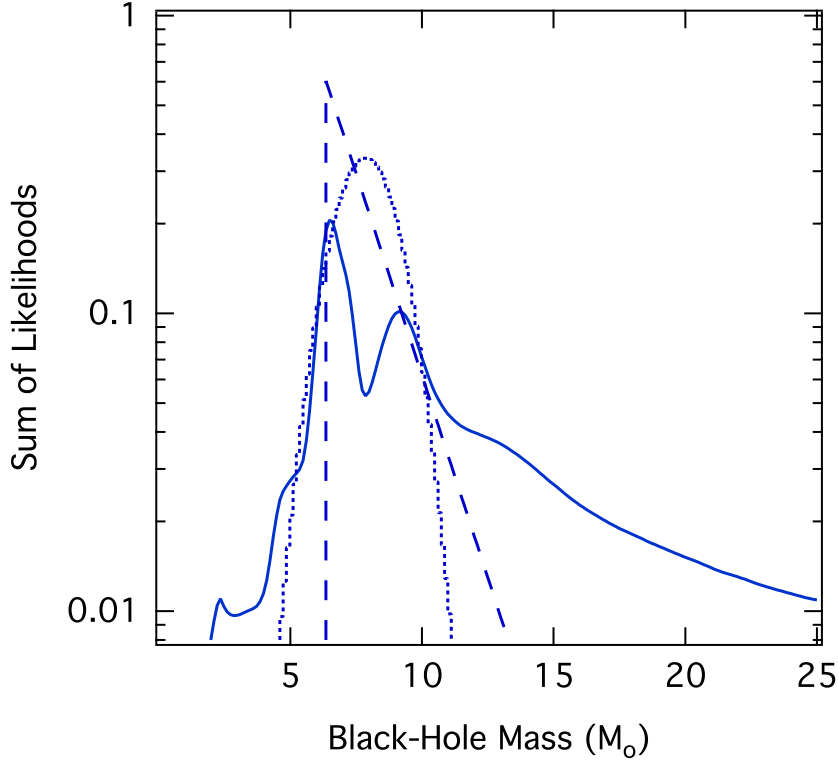}
\caption{The solid line shows the sum of likelihoods for the mass 
measurements of the 16 black holes in low-mass X-ray binaries. Note
that because of the high-mass wings of the individual likelihoods, the
shape of their sum is artificial at the high mass end. The dashed and
dotted lines show the exponential and Gaussian distributions,
respectively, with parameters that best fit the data (see \S4).}
\label{fig:prob_added} 
\end{figure}

A clustering of the observed black hole masses between $\sim
6-10~M_{\odot}$ is already evident from Figure~\ref{fig:bh_prob}. In
the next section, we will carry out a formal Bayesian analysis to
determine the parameters of the underlying mass distribution that is
consistent with the observed $P_i({\rm data}\vert M)$ shown here. If
the likelihood for each source was narrow enough such that there was
little or no overlap between them, then adding the likelihoods for the
entire sample and coarsely binning the resulting distribution would
provide a good estimate of the underlying mass distribution. Even
though this condition is not entirely satisfied here, especially at
the high mass end, we nevertheless show in Figure~\ref{fig:prob_added}
this approximate mass distribution to get a sense of its gross
properties.
 
%%%%%%%%%%%%%%%%%%%%%%
%%%%   Table 2  %%%%%%
%%%%%%%%%%%%%%%%%%%%%%

\begin{deluxetable}{llllllll}
\tabletypesize{\scriptsize}
%\rotate
\tablewidth{0pt}
\tablenum{2}
\tablecaption{Dynamical Data for Twenty-Three Black Hole Binaries}
\tablehead{
 \colhead{} &
 \colhead{Coordinate} &
 \colhead{Common Name} &
 \colhead{$f(M)$} &
 \colhead{$q$} &
 \colhead{$i$} &
 \colhead{$M$} &
 \colhead{References\tablenotemark{b}}
\cr
 \colhead{} &
 \colhead{Name} &
 \colhead{or Prefix\tablenotemark{a}} &
 \colhead{($M_{\odot}$)} &
 \colhead{$q_{\rm min} - q_{\rm max}$} &
 \colhead{(deg)} &
 \colhead{($M_{\odot}$)} &
 \colhead{} 
}
\startdata
% Table: Dynamical Data for 23 Black Hole Binaries
%
%   Coordinate    Common Name  f(M)                 q            i                M               Ref.     
%   Name          or Prefix    (Msun)          qmin--qmax       deg             (Msun)           (Mass)   
%--------------------------------------------------------------------------------------------------------------------

1&  0620-003&     (A)&         $2.76\pm0.01$&  0.056--0.064&    $51.0\pm0.9$&    $6.6\pm0.25$&    1,2\\
2&  1543-47&      (4U)&        $0.25\pm0.01$&  0.25--0.31&      $20.7\pm1.5$&    $9.4\pm1.0$&     3\\
3&  1550-564&     (XTEJ)&      $7.73\pm0.40$&  0.0--0.040&      $74.7\pm3.8$&    $9.1\pm0.6$&     4\\
4&  1655-40&      (GROJ)&      $2.73\pm0.09$&  0.37--0.42&      $70.2\pm1.9$&    $6.3\pm0.27$&    3,5\\
5&  1819.3-2525&  V4641 Sgr&   $3.13\pm0.13$&  0.42--0.45&      $75\pm2$&        $7.1\pm0.3$&     3\\
6&  2023+338&     (GS)&        $6.08\pm0.06$&  0.056--0.063&    $55\pm4$&        $12\pm2$&        6\\

\hline

7&  0422+32&      (GROJ)&      $1.19\pm0.02$&  0.076--0.31&     \nodata&         \nodata&         3,7\\
8&  1009-45&      (GRS)&       $3.17\pm0.12$&  0.12--0.16&      \nodata&         \nodata&         3,8\\
9&  1118+480&     (XTEJ)&      $6.1\pm0.3$&    0.035--0.044&    \nodata&         \nodata&         3,9,10\\
10& 1124-683&     Nova Mus 91& $3.01\pm0.15$&  0.11--0.21&      \nodata&         \nodata&         3,11\\
11& 1354-64&      (GS)&        $5.73\pm0.29$&  0.08--0.15&      \nodata&         \nodata&         12\\
12& 1650-500&     (XTEJ)&      $2.73\pm0.56$&  0.0--0.2\tablenotemark{c}&        \nodata&    \nodata&     13\\
13& 1659-487&     GX 339-4&    $5.8\pm0.5$&    0.0--0.4\tablenotemark{c}&        \nodata&    \nodata&     14,15\\
14& 1705-250&     Nova Oph 77& $4.86\pm0.13$&  0.0--0.053&      \nodata&         \nodata&         3,6\\
15& 2000+251&     (GS)&        $5.01\pm0.12$&  0.035--0.053&    \nodata&         \nodata&         3,6\\
16& 1915+105&     (GRS)&       $9.5\pm3.0$&    0.025--0.091&      $66\pm$2&      \nodata&         16,17,18,19\\
17& 1859+226&     (XTEJ)&      $7.4\pm1.1$\tablenotemark{d}&  \nodata& \nodata&  \nodata&         20,21\\

\hline

18& 0020+593&     IC 10 X-1&   $7.64\pm1.26$&   \nodata&        \nodata&         $>20$&           22,23\\
19& 0055-377&     NGC 300-1&   $2.6\pm0.3$&     \nodata&        \nodata&         $>10$&           24\\
20& 0133+305&     M33 X-7&     $0.46\pm0.07$&   \nodata&        \nodata&         $15.65\pm1.45$&  25\\
21& 0538-641&     LMC X-3&     $2.3\pm0.3$&     \nodata&        \nodata&         \nodata&         26\\
22& 0540-697&     LMC X-1&     $0.886\pm0.037$& \nodata&        \nodata&         $10.91\pm1.54$&  27\\
23& 1956+350&     Cyg X-1&     $0.251\pm0.007$& \nodata&        \nodata&         $>8$&            28,29

\enddata
\scriptsize

\tablenotetext{a} {The entries in parentheses are prefixes to the
  coordinate names that identify the discovery X-ray mission.}
\tablenotetext{b} {References: 
1.\ Neilsen et al.\ 2008; 2.\ Cantrell et al.\ 2010;
3.\ Orosz 2003; 
4.\ Orosz et al.\ 2010; 
5.\ Greene et al.\ 2001; 
6.\ Charles \& Coe 2006;
7.\ Filippenko et al.\ 1995;
8.\ Filippenko et al.\ 1999;
9.\ McClintock et al.\ 2001; 10.\ Gelino et al.\ 2006;
11.\ Orosz et al.\ 1996;
12.\ Casares et al.\ 2009; 
13.\ Orosz et al.\ 2004;
14.\ Hynes et al.\ 2003; 
15.\ Mu{\~n}oz-Darias et al.\ 2005; 
16.\ Greiner et al.\ 2001; 17.\ Neil et al.\ 2007; 18.\ Harlaftis \& Greiner 2004; 19. Fender et al.\ 1999;
20.\ Filippenko \& Chornock 2001; 21.\ Zurita et al.\ 2002;
22.\ Prestwich et al.\ 2007; 23.\ Silverman \& Filippenko 2008; 
24.\ Crowther et al.\ 2010; 
25.\ Orosz et al.\ 2007; 
26.\ Cowley 1992; 
27.\ Orosz et al. 2009; 
28.\ Caballero-Nieves et al.\ 2009; 29.\ M.\ Reid, private communication.}
\tablenotetext{c} {Estimated range based on extreme values observed for systems with comparable orbital periods.}
\tablenotetext{d} {Unconfirmed and uncertain value; see text.}
\end{deluxetable}

%%%%%%%%%%%%%%%%%%%%%%%%%%%%%%%%%%%%%%%%%%%%%%%%%%%%%%%%%%%%%%%%%%%%%%

\section{The Intrinsic Distribution of Black Hole Masses}

In this section, we use a parametric form of the black-hole mass
distribution and the data discussed in Sections 2 and 3 in order to
determine its parameters. We will first consider an exponentially
decaying mass distribution with a cut-off given by
\begin{equation}
P(M; M_{\rm scale}, M_{\rm c})=\frac{\exp(M_{\rm c}/M_{\rm scale})}{M_{\rm scale}} 
\left\{\begin{array}{ll}
\exp(-M/M_{\rm scale})\;, & M>M_{\rm c}\\
0\;, & M\le M_{\rm c}
\end{array}
\right.\;.
\label{eq:mass_distr}
\end{equation}
This choice of the mass distribution is motivated by theoretical
expectations based on the energetics of supernova explosions, as well
as the density profiles and mass distributions of pre-supernova
stars. The typical value of the mass scale is expected to lie in the
range $M_{\rm scale} \sim 5.5 - 9~M_\odot$ (as we infer from the
various figures in Fryer \& Kalogera 2001), whereas the cutoff mass is
simply expected to be the maximum neutron star mass. Our goal is to
find the values of the mass scale $M_{\rm scale}$ in the exponential
and the cut-off mass $M_{\rm c}$ that maximize a properly defined
likelihood and to estimate their uncertainties. We will show below
that the particular choice of the functional form of the mass
distribution does not affect the main conclusions of the paper.

In Section~3, we calculated, for each observed black hole, the
probability $P_i({\rm data}\vert M)$, which measures the chance of
making a particular observation if the black hole has mass $M$.  What
we want to calculate here is the probability $P(M_{\rm scale},M_{\rm
c}\vert {\rm data})$, which measures the likelihood of the parameters
of the black hole mass distribution, given the observations. Using
Bayes' theorem, we can write this as
\begin{equation} 
P(M_{\rm scale},M_{\rm c}\vert {\rm data}) = C_2 P({\rm data}\vert M_{\rm scale}, M_{\rm c}) 
P(M_{\rm scale})P(M_{\rm c})\;,
\label{eq:bayes} 
\end{equation} 
where $C_2$ is the normalization constant and $P(M_{\rm scale})$ and $P(M_{\rm c})$
are the priors over the values of the mass scale and cut-off
mass. We assume a flat prior over
the mass scale between $M_{\rm scale}=0$ and a maximum value $M_{\rm scale}=M_{\rm max}$,
i.e., 
\begin{equation} 
P(M_{\rm scale})=\left\{\begin{array}{ll} 0, & M_{\rm scale}\le 0\\
\frac{1}{M_{\rm max}}, & 0<M_{\rm scale} \leq M_{\rm max}\\ 
0, & M_{\rm scale}>M_{\rm max}\;.
\end{array}\right.
\label{eq:likely} 
\end{equation} 
The upper limit $M_{\rm max}$ is imposed mostly for computational
reasons and does not affect the results. We also adopt a similar prior
over the cut-off mass between the maximum neutron-star mass, which we
set to 2~$M_\odot$, and the minimum well-established mass measurement
for a black hole. As will be evident from the results, the particular
choice of this range does not affect the measured parameters.  We also
repeated this analysis with logarithmic priors in the two parameters
and found that the results are insensistive to the choice of priors.

In equation~(\ref{eq:bayes}), the quantity $P({\rm data}\vert M_{\rm
scale}, M_{\rm c})$ measures the chance that we make a particular set
of observations for the ensemble of black holes, given the values of
the parameters of the mass distribution. We need now to estimate this
quantity, given the likelihoods for the individual sources. We will
assume that each measurement is independent of all the others, so that
\begin{equation}
P({\rm data}\vert M_{\rm scale}, M_{\rm c})=
\prod_i \int dM P_i ({\rm data}| M) P(M;M_{\rm scale}, M_{\rm c})\;.
\end{equation}
Combining this last equation with equation~(\ref{eq:bayes}) we obtain
\begin{equation}
P(M_{\rm scale},M_{\rm c}\vert {\rm data})=C
P(M_{\rm scale})P(M_{\rm c})
\prod_i \int dM P_i ({\rm data}| M) P(M;M_{\rm scale}, M_{\rm c})\;, 
\label{eq:final}
\end{equation}
where $C$ is the overall normalization constant. 

\begin{figure}
\centering
   \includegraphics[scale=0.75]{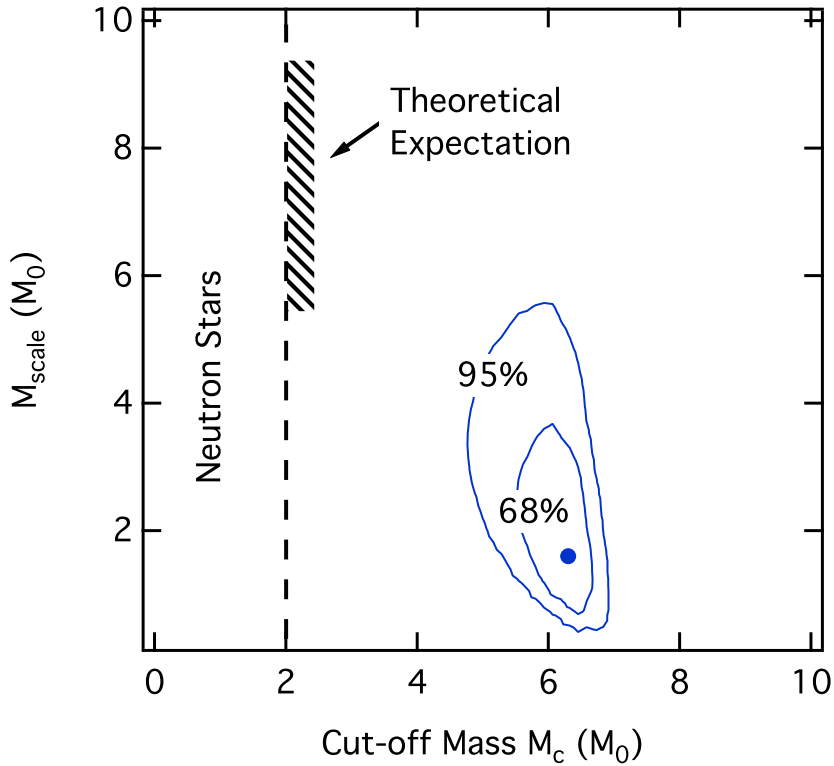}
\caption{The parameters of an exponential black hole mass distribution 
with a low-mass cutoff. The cut-off mass is well above theoretical
expectations, indicating a sizable gap between neutron-star and black
hole masses. Furthermore, the mass scale in the exponential is
significantly smaller than theoretical expectations. }
\label{fig:bayes_exp} 
\end{figure}

We show in Figure~\ref{fig:bayes_exp} the 68\% and 95\% confidence
contours of the mass scale and cut-off mass that best describe the
observations and compare them to the theoretical expectation. The lack
of black holes below $\sim 5~M_\odot$ and the rapid decline of the
exponential distribution at the high mass end are both remarkable (see
the dashed line in Fig.~\ref{fig:prob_added}). The latter result is
not at odds with the relatively high mass of GRS~1915+105 because of
the wide and shallow mass probability distribution of this source.

\begin{figure}
\centering
   \includegraphics[scale=0.75]{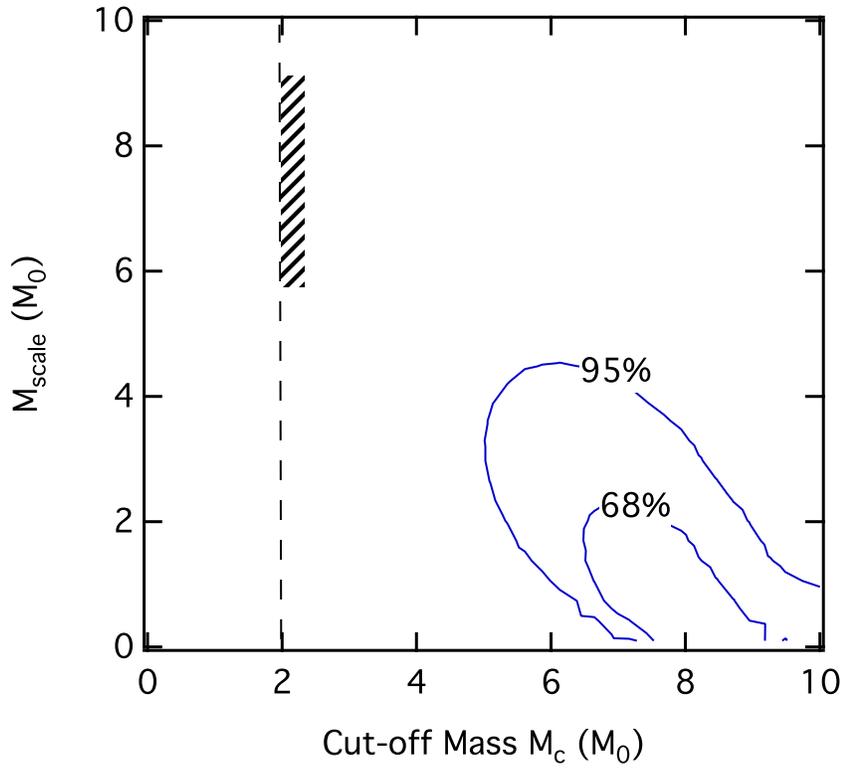}
\caption{The parameters of an exponential black-hole mass distribution 
when all possible inclinations for each binary system were allowed
(i.e., no inclination information was taken into account).  The
existence of the low-mass gap and the small exponential mass scale
remain robust.}
\label{fig:no_inc} 
\end{figure}

In order to explore whether the small number of sources with very
well-determined masses dominate this result, we did the following
test.  We repeated the calculation using the mass functions and
constraints on the mass ratios for all sources but ignoring any
information on the inclinations of the binary systems. This, in
effect, is equivalent to treating all sources in category {\it (i)}
using the formalism we applied to sources in category {\it (ii)},
integrating over all possible values of inclination.
Figure~\ref{fig:no_inc} shows the 68\% and 95\% confidence contours of
the parameters of the exponential distribution in this test
case. Although the allowed range of values is increased, as expected,
the low-mass gap and the discrepancy with the theoretically expected
mass scale remain robust.

\begin{figure}
\centering
   \includegraphics[scale=0.75]{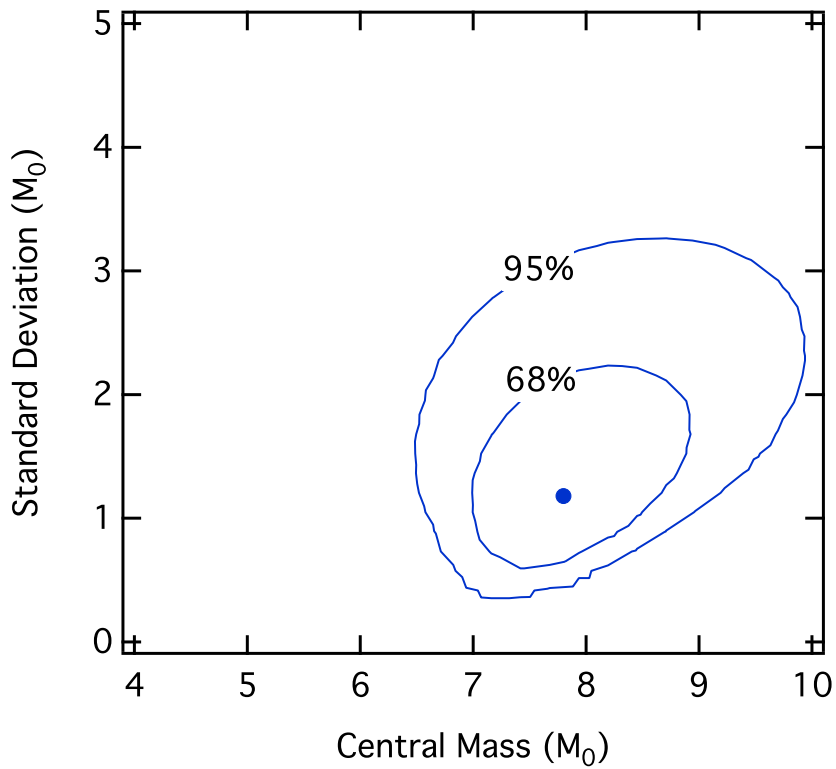}
\caption{The parameters of a Gaussian black hole mass distribution. 
The data are best described by a narrow distribution at $7.8 \pm 
1.2~M_\odot$. }
\label{fig:bayes_gauss} 
\end{figure}

The narrowness of the mass distribution implied by the above results
motivated us to explore different functional forms of the underlying
distribution, and in particular, a Gaussian function. The Gaussian
function here serves as a phenomenological two-parameter description
of a narrow distribution and is not necessarily motivated by
theory. We show in Figure~\ref{fig:bayes_gauss} the parameters of such
a Gaussian distribution that best describes the observations. The
masses of all 16 black holes are consistent with a narrow distribution
at $7.8 \pm 1.2~M_\odot$. This result is in agreement with an earlier,
more limited, study by Bailyn et al. (1998).

%%%%%%%%%%%%%%%%%%%%%%%%%%%%%%%%%%%%%%%%%%%%
%%%%%%%%%%%   Table 3  %%%%%%%%%%%%%%%%%%%%%
%%%%%%%%%%%%%%%%%%%%%%%%%%%%%%%%%%%%%%%%%%%%

\begin{deluxetable}{llllllll}
\tabletypesize{\scriptsize}
%\rotate
\tablewidth{0pt}
\tablenum{3}
\tablecaption{Thirty-Two Transient Black-Hole Candidates}
\tablehead{
 \colhead{} &
 \colhead{Source Name} &
 \colhead{RA(J2000)} &
 \colhead{Dec(J2000)} &
 \colhead{$l$(deg)} &
 \colhead{$b$(deg)} &
 \colhead{I(Crab)\tablenotemark{a}} &
 \colhead{References\tablenotemark{b}}
}
\startdata

%    Name                  RA(J2000)    Dec(J2000)       l(deg)    b(deg)      Max.\ Int.  References
%-----------------------------------------------------------------------------------------------------
1&   1A 1524-61&           15 28 17.2&   -61 52 58&      320.3&    -4.4&   0.95&   1\\

2&   4U 1630-47&           16 34 01.61&  -47 23 34.8&    336.9&    +0.3&   1.4&    1\\

3&   XTE J1652-453&        16 52 20.33&  -45 20 39.6&    340.5&    -0.8&   0.19&   2\\

4&   IGR J17098-3628&      17 09 45.93&  -36 27 58.2&    349.6&    +2.1&   0.13&   1\\  

5&   SAX J1711.6-3808&     17 11 37.1&   -38 07 05.7&    348.6&    +0.8&   0.13&   1\\

6&   GRO J1719-24&         17 19 36.93&  -25 01 03.4&    359.9&    +7.0&   1.5&    1\\

7&   XTE J1720-318&        17 19 58.994& -31 45 01.25&   354.6&    +3.1&   0.41&   1\\

8&   IGR J17269-4737&      17 26 49.28&  -47 38 24.9&    342.2&    -6.9&   0.083&  1\\

9&   GRS 1730-312&         17 33 52.3&   -31 12 25&      356.7&    +1.0&   0.20&   1\\     

10&  GRS 1737-31&          17 40 09&     -31 02 24&      357.3&    +0.6&   0.026&  1\\

11&  GRS 1739-278&         17 42 40.03&  -27 44 52.7&    0.7&      +1.2&   0.86&   1\\

12&  KS J1739-304&         17 42 44.0&   -30 30 51&      358.3&    -0.3&   0.009&  1\\   

13&  1E 1740.7-2942&       17 43 54.83&  -29 44 42.6&    359.1&    -0.1&   0.03&   1\\

14&  1A 1742-289&          17 45 37.0&   -29 01 07&      359.9&    -0.0&   2.0&    1\\    

15&  H1743-322&            17 46 15.57&  -32 14 01.1&    357.1&    -1.6&   0.77&   1\\

16&  XTE J1748-288&        17 48 05.06&  -28 28 25.8&    0.7&      -0.2&   0.64&   1\\

17&  IGR J17497-2821&      17 49 38.037& -28 21 17.37&   1.0&      -0.5&   0.03&   1\\

18&  SLX 1746-331&         17 49 48.3&   -33 12 26&      356.9&    -3.1&   0.27&   1\\        

19&  Swift J1752-223&      17 52 15.10&  -22 20 32.78&   6.4&      +2.1&   0.11&   3\\

20&  Swift J1753.5-0127&   17 53 28.29&  -01 27 06.22&   24.9&     +12.2&  0.16&   4\\

21&  XTE J1755-324&        17 55 28.6&   -32 28 39&      358.0&    -3.6&   0.18&   1\\ 

22&  4U 1755-33&           17 58 40.0&   -33 48 27&      357.2&    -4.9&   0.10&   1\\

23&  GRS 1758-258&         18 01 12.40&  -25 44 36.1&    4.5&      -1.4&   0.02&   1\\

24&  SAX J1805.5-2031&     18 05 34&     -20 30 48&      9.6&      +0.3&   0.051&  1\\

25&  XTE J1817-330&        18 17 43.54&  -33 01 07.8&    359.8&    -8.0&   1.35&   1\\

26&  XTE J1818-245&        18 18 24.4&   -24 32 18.0&    7.4&      -4.2&   0.51&   1\\

27&  Swift J1842.5-1124&   18 42 17.33&  -11 25 00.6&    21.7&     -3.2&   0.04&   5\\

28&  EXO 1846-031&         18 46 39.8&   -03 07 21&      29.9&     -0.9&   0.3&    1\\

29&  IGR J18539+0727&      18 53 54&     +07 27&         39.8&     +2.8&   0.02&   1\\

30&  XTE J1856+053&        18 56 39&     +05 19 48&      38.3&     +1.3&   0.07&   1\\      

31&  XTE J1908+094&        19 08 53.077& +09 23 04.90&   43.3&     +0.4&   0.10&   1\\

32&  XTE J2012+381&        20 12 37.71&  +38 11 01.1&    75.4&     +2.2&   0.16&   1

\enddata
\scriptsize

\tablenotetext{a}{Approximate maximum X-ray intensity from the Liu et al.\ (2007) catalogue 
supplemented in a few cases by data from other sources.  In some instances, the intensity is for 
a higher enery band than the standard $\sim2-10$~keV band.}

\tablenotetext{b}{References:
1.\ Liu et al.\ 2007;
2.\ Markwardt et al.\ 2009;
3.\ Torres et al.\ 2009;
4.\ Fender et al.\ 2005;
5.\ Krimm et al.\ 2008.}
\end{deluxetable}

%%%%%%%%%%%%%%%%%%%%%%%%%%%%%%%%%%%%%%%%%%%%%%%%%%%%%%%%%%%%%%%%%%%%%%

\section{Observational Selection Effects}

In the previous section, we discussed two distinct results we obtained
on the mass distribution of stellar black holes using the data on 16
sources in low-mass X-ray binaries: The cut-off mass at the low end is
$\gtrsim 5~M_\odot$, indicating a significant lack of black holes in
the $\sim 2-5~M_\odot$ range, and the mass scale in the exponential is
$\sim 1.2~M_\odot$, indicating a rapid decline of the distribution at
the high-mass end. In other words, the black hole mass distribution is
narrow and can also be represented with a Gaussian at $7.8 \pm
1.2~M_\odot$. We now explore two questions about our results: {\em
(i)\/} whether any observational selection effects could have biased
the black-hole mass distribution we inferred for transient sources and
{\em (ii)\/} whether the black holes in transient low-mass X-ray
binaries are a representative sample of the population of stellar
black holes in the Galaxy.

\subsection{Selection Biases in the Mass Distribution of Black Holes in
Transient X-ray Binaries}

The masses of stellar black holes have generally been measured when
the black holes {\it (i)\/} reside in binary systems, {\it (ii)\/} are
transient X-ray sources, and {\it (iii)\/} reach peak fluxes above
1~Crab (about $2.5 \times 10^{-8}$~erg~s$^{-1}$~cm$^{-2}$) during
their outbursts. The first requirement naturally arises from the
nature of the dynamical mass measurements. The second requirement
ensures that optical observations of the companion star can be carried
out during the quiescent phase so that the radial velocity
measurements are not contaminated by the emission from the accretion
disk\footnote{The dynamical Bowen technique (Steeghs \& Casares 2002)
offers the possibility of measuring the masses of black holes in
persistent systems.}. Finally, the third requirement is a common
strategy of observing campaigns that aims to select among the
transients those sources that are the closest and the least obscured.

\begin{figure}
\centering
   \includegraphics[scale=0.75]{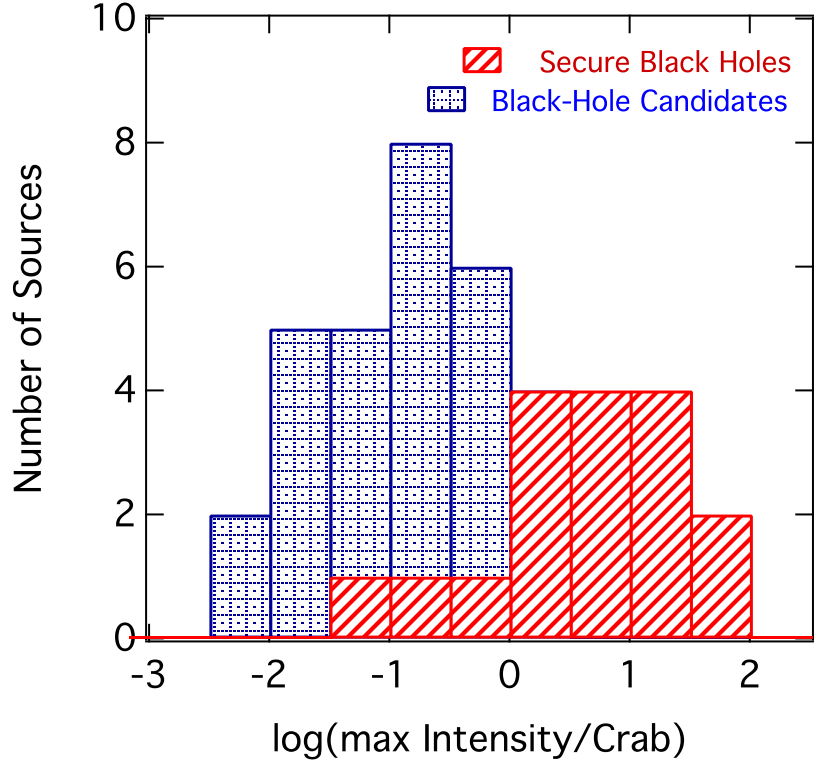}
\caption{The histogram of maximum flux observed during the outbursts 
of black holes with dynamical mass determinations (red) and other
black hole candidates (blue). A clear dichotomy between the two groups
at 1~Crab is evident.}
\label{fig:hist} 
\end{figure}

Black hole transients have been observed in the last four decades with
a large collection of X-ray telescopes. We, therefore, tried to verify
and quantify the third requirement for follow-up by compiling the
highest flux level observed from each transient black hole or black
hole candidate in outburst by any instrument (see Table~1). We show in
Figure~\ref{fig:hist} the histogram of the highest flux levels
observed during outburst from the 16 confirmed black holes (in red)
and from the black hole candidates (in blue). The dichotomy between
the fluxes of the two groups is striking and indeed shows that
primarily the brightest sources have been followed up for dynamical
mass measurements.

\begin{figure}
\centering
   \includegraphics[scale=0.75]{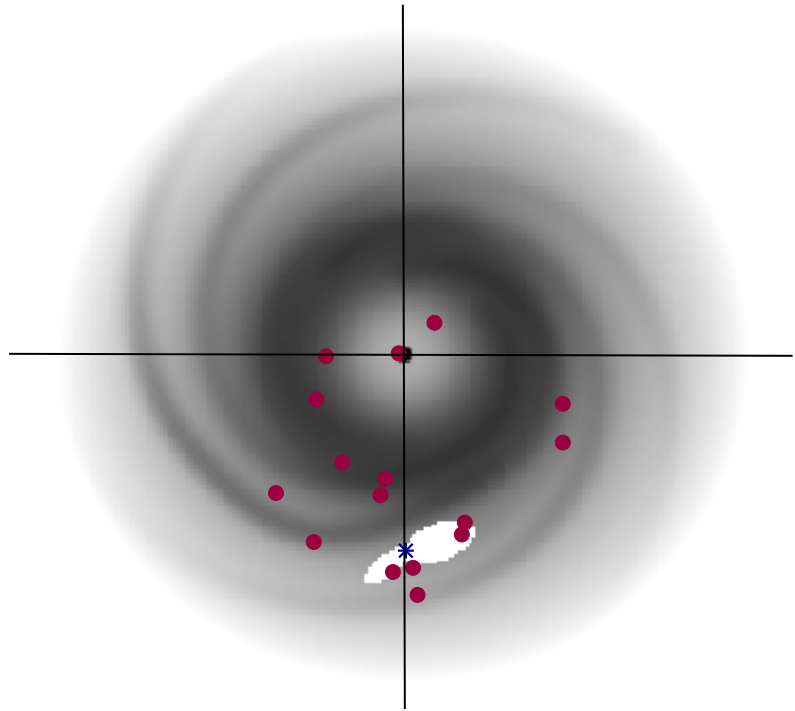}
\caption{The positions on the Galactic disk of the black holes with secure 
dynamical mass measurements. The bakground grayscale image is the
electron distribution in the Galaxy according to the model of Cordes
\& Lazio (2001). The plus sign represents the position of the Sun.}
\label{fig:gal_distr} 
\end{figure}

We then explored whether the 1~Crab flux threshold during outburst
could lead to a selection bias against low-mass black holes. Setting a
flux limit for optical follow-ups restricts, in principle, the volume
of the Galaxy that is sampled. If, in addition, the low-mass black
holes form an intrinsically fainter population than that of more
massive black holes, this could indeed introduce a selection bias.

Consider, for example, the case in which the maximum luminosity during
the outburst of an X-ray transient is proportional to the mass of the
black hole, so that the binary systems with low-mass black holes are
intrinsically dimmer than their high mass counterparts.  Because the
volume of the disk component of the Galaxy scales as the distance
squared, the volume sampled by sources selected for follow-up
observations scales as the mass of the black hole. If the mass
distribution of black holes was rather flat, then, on average, $\sim
4$ times more $\sim 8~M_\odot$ black holes would be followed up than
$\sim 2~M_\odot$ black holes, leading to an artificial reduction in
the number of low-mass black holes.

If there were indeed a population of intrinsically fainter, low-mass
black holes, for the 16 more massive ones that have been observed, 4
low mass black holes should have been found within the same flux
limits according to the previous argument. In contrast, there is none
in the current sample. This fact argues that the lack of low-mass
black holes is real.

Another way to determine whether the volume-limited sample introduces
a mass bias is to compare the number and distribution in the Galaxy of
secure black hole sources to those of the black hole candidates.  We
plot in Figure~\ref{fig:gal_distr} the galactic distribution of the
black holes with secure mass determinations. It is evident that these
sources lie within $\simeq 1/3$ of the volume of the galactic disk. As
a result, if the observed black holes are a representative sample of
the true population, our catalogue of black-hole candidates should
contain twice as many sources that are fainter simply because they lie
further away in the Galaxy compared to the 16 sources that have been
followed up. This is precisely the ratio of the number of black-hole
candidate sources in Table~3 to that of secure black holes in Table~1:
32:16. 

The 16 secure black holes have a median intensity of 3.6~Crab and a
median distance of $\sim$4~kpc.  The 32 black hole candidates, on the
other hand, have a median intensity of 0.18~Crab, so we expect them to
be on average $\sqrt{3.6/0.18}=4.5$ times farther. (Note, however,
that confirmed black holes have benefited from almost continuous
monitoring that led to an accurate measurement of their the maximum
intensities. The black hole candidates have been observed more
sparsely, so that the maximum reported intensity is in most cases
likely to be below the actual maximum intensity that the sources
reached). This further supports the idea that the black hole
candidates lie in the far two-thirds of the Galaxy.

We, therefore, conclude that there is no room within the 32 black-hole
candidates to hide any sizable population of sources that are fainter
because they are intrinsically less massive.  The fact that our sample
of secure black-holes is flux limited does not imply that it is
luminosity limited, and, therefore, does not introduce any selection
bias against the detection of low-mass black holes.

\subsection{Are Black Holes in Transient Low-Mass X-ray Binaries a 
Representative Sample?}

The sample of 16 sources that we used to infer the mass distribution
of stellar black holes has been drawn from transient low-mass X-ray
binaries\footnote{Persistent sources with O/B-type secondaries have
also been successfully targeted for mass measurements; see Tables~1
and 2.}. Here, we consider the possibility that low-mass black holes
exist, but not in transient low-mass X-ray binaries. In turn, we will
consider the possibility that low-mass black holes are primarily in
persistent binaries or that they are not members of binary systems at
all.

\smallskip

\noindent {\em (i) Are accreting low-mass ($2-5~M_\odot$) or high mass 
($\gtrsim 9~M_\odot$) black holes persistent X-ray sources?---\/} Most
known neutron-star low-mass X-ray binaries are persistent sources,
whereas the majority of black hole binaries are transient
systems. (The latter statement is primarily a consequence of the fact
that the transient behavior has so far been required in establishing
the black hole nature of a compact object in a majority of the
sources). Is it possible that low-mass black holes behave more like
neutron stars than their more massive counterparts?

The most prominent explanation of the transient behavior of X-ray
binaries is the irradiated disk instability model (van Paradijs 1996;
King, Kolb, \& Burderi 1996; Dubus et al.\ 1999). In this model,
whether an X-ray binary is a transient depends only weakly on the
nature of the primary but is mainly determined by the orbital period
of the system, the mass of the primary, and the evolutionary state of
the companion star (King et al.\ 1996). Thus, the expectation is that
a 4~$M_\odot$ black hole binary should behave like one containing a
6~$M_\odot$ black hole if they have comparable orbital
periods. Furthermore, our sample of 16 transient black hole sources
includes both short-period and long-period systems, indicating that
the transient behavior is seen over a range of orbital periods. We,
therefore, find it unlikely, albeit still possible, that low-mass
black holes would preferentially exist in persistent binaries if the
distribution of their orbital periods is similar to that of the
confirmed black holes. 

We, nevertheless, explore whether the known number of persistent X-ray
sources in the Galaxy is large enough to harbor the low-mass back
holes if, indeed, all of the latter are persistent systems. To this
end, we analyzed the catalogue of low-mass X-ray binaries by Liu, van
den Heuvel, \& van Paradijs (2007).  Starting with the total number of
non-pulsing X-ray binaries in the Galaxy, we identified the following
categories: {\it (a)\/} Confirmed black holes, which refer to those
with dynamical mass measurements; {\it (b)\/} Black hole candidates,
which are sources with no mass determinations, but possess spectra and
radio properties that resemble those of the confirmed black holes;
{\it (c)\/} Unidentified sources, which are those with significant
number of observations but no distinguishing characteristics to reveal
the nature of the primary; {\it (d)\/} Neutron stars, which are
sources that exhibit thermonuclear X-ray bursts; and {\it (e)\/}
Sources for which there is very little information. In Table~4, we
show the numbers and relative fractions of sources in these
categories.

Table~4 indicates that there are a total of 72 known persistent X-ray
sources. 46 out of these 72 sources have shown thermonuclear bursts
and can securely be identified as neutron stars, leaving at most 26
sources as possible low-mass black hole binaries. In reality, only two
of these remaining 26 sources show evidence for black hole nature, and
a further 7 show no thermonuclear bursts even though they have been
extensively observed. We will, nevertheless, allow the possibility
that all can harbor low-mass black holes.

For any integrable (i.e., declining faster than logarithmic)
distribution of black hole masses that does not have a low-mass gap,
there should be more black holes with masses in the $2-5 M_\odot$
range than with masses $\ge 5 M_\odot$. If all the 26 persistent
sourcse were low-mass black holes, we would expect this number to be
larger than the total number of transient (higher mass) black hole
sources in the Galaxy. From the galactic distribution of transient
black-hole binaries, we estimated (see previous subsection) that there
are at least three times as many black holes in the Galaxy with masses
$\ge 5 M_\odot$ than our current sample of 16.  Taking into account
the short duty cycles of transient black-hole binaries, we can safely
infer that the total number of such systems in the galaxy is $\gg 48$,
which is significantly larger than 26. As a result, even if all the 26
non-bursting sources were low-mass black holes, they would not be
enough to efficiently close the low-mass gap between $\sim 2$ and
$\sim 5~M_\odot$ in the inferred distribution. Nevertheless, applying
the dynamical Bowen technique to persistent low-mass X-ray binaries
will help explore further this possibility.

\smallskip

\noindent {\em (ii) Are low-mass ($2-5~M_\odot$) or high mass 
($\gtrsim 9 M_\odot$) black holes not members of X-ray binary
systems?---\/} The fraction of binaries that survive the supernova
explosion is determined by a number of factors, such as the orbital
separation, the amount of mass loss and the kick velocity during the
supernova explosion. These factors depend on the progenitor and the
remnant masses and may work in a way to hinder the formation of
low-mass X-ray binaries with low- or high-mass black holes. Moreover,
even if the binaries survive the explosions, systems with low- or
high-mass remnants may become wider and perhaps never reach contact.
It appears unlikely that this is the explanation of the paucity of
low-mass black holes between the systems with dynamical mass
measurements (see Fryer \& Kalogera 2001), although addressing this
question would only be possible through detailed population synthesis
calculations.

On the other hand, the particular evolutionary path that leads to the
formation of black-hole X-ray binaries might be responsible for the
lack of systems with relatively massive black holes ($\gtrsim
9~M_\odot$). Mass transfer in the low-mass X-ray binary phase requires
small orbital separations between the black hole and the low-mass
companion. The common envelope phase, which is invoked to reduce
sufficiently the orbital separation of the binary before the
supernova, is thought to lead to the expulsion of the hydrogen
envelope of the pre-supernova star, leaving behind a bare helium
core. Furthermore, the winds from the resulting helium cores in a
Wolf-Rayet phase are expected to lead to further mass loss, albeit at
rates that are highly uncertain. As a result, it is possible that the
black hole masses in contact binaries are capped at $\simeq
10~M_\odot$, although this value depends strongly on the assumed mass
loss rates (Woosley, Heger, \& Weaver 2002). This evolutionary path
could provide a natural explanation for the rapid decline of the
inferred mass function at the high-mass end. The same mechanism does
not necessarily cap the mass of the black holes in high-mass X-ray
binaries, which are wind-fed.

\section{Implications for Black Hole Formation}

We now explore the implications of the narrow mass distribution of
black holes in low-mass X-ray binaries for the progenitors of these
systems and the supernovae that form the black holes.

As we discussed in the previous section, the rapid decline at the high
mass end of the inferred distribution is plausibly the result of the
particular binary formation channel that leads to low-mass X-ray
binaries. In contrast to this population, relatively more massive
black holes ($> 20~M_\odot$) have been detected in high mass X-ray
binaries (see the persistent sources in Table~2). This argues that
there is no intrinsic barrier for the formation of massive black holes
as the end stages of stellar evolution but that the differences
between the black hole masses in the two binary populations are due to
the different evolutionary paths that they follow. Observational
studies of mass loss from stars with hydrogen and helium envelopes, in
isolation and in tight binary systems, will elucidate and test these
ideas.

The low-mass gap, on the other hand, cannot be accounted for by
observational selection effects (even though the mass measurement
method requires a particular choice of targets and the observing
strategy has resulted in a flux-limited sample). It is also unlikely
that it results from the particular evolutionary paths of low-mass
X-ray binaries. Therefore, it appears to be real.

The observed distribution of masses could be different than the
distribution of birth masses because mass transfer in the X-ray binary
phase could have caused the black holes to evolve away from the masses
they were born with. Population synthesis models (e.g., Fragos et al.\
2010) show, however, that the least massive black holes are
preferentially born with low-mass companions ($\lesssim 2 M_\odot$)
and, therefore, could not accrete a significant amount of matter over
the lifetime of the binary. When the companions are more massive than
5~$M_\odot$ at the start of the mass transfer phase, the black holes
are born with masses that are also $\gtrsim 5~M_\odot$. As a result,
the accretion in the binary phase is not likely to bridge the gap
between 2 and 5~$M_\odot$.

It has been suggested that black hole evaporation in braneworld
gravity models can lead to a deficiency of low-mass black holes in the
steady state population. A gap could be created in this context
because the rate of evaporation in braneworld gravity is very rapid
and increases with decreasing black hole mass (Postnov \&
Cherepashchuk 2003). However, recent constraints on the rate of
evaporation obtained using the current population of black holes
preclude this possibility (Johannsen, Psaltis, \& McClintock 2009).

The paucity of black holes with masses less than $5~M_\odot$ are
likely, therefore, to be related to the physics of supernova
explosions that lead to the formation of black holes. Numerical
simulations of supernova explosions typically generate a continuous
distribution of black hole masses that decays as an exponential (Fryer
1999; Fryer \& Kalogera 2001). The continuity of masses is primarily a
consequence of the relatively gradual dependence of explosion energies
on the masses of the progenitors and the fact that explosion energies
are still comparable to (although smaller than) the binding energy of
the stellar envelopes.  Although creating a mass gap is difficult
given the current understanding of the supernova energetics, it has
been suggested that it can be achieved under the ad hoc assumption
that the explosion energy has a step-function dependence on progenitor
mass and that it plunges to zero for stars more massive than $\sim
25~M_\odot$ (Fryer \& Kalogera 2001). 

Such a bimodality in the energies of explosions that form neutron
stars versus black holes should become apparent in the large sample of
supernovae anticipated from the ongoing surveys that are sensitive to
underluminous core-collapse supernovae.

%%%%%%%%%%%%%%%%%%%%%%%%%%%%%%%%%%%%%%%%%%%%
%%%%%%%%%%%   Table 4  %%%%%%%%%%%%%%%%%%%%%
%%%%%%%%%%%%%%%%%%%%%%%%%%%%%%%%%%%%%%%%%%%%

\begin{deluxetable}{cccc}
\tabletypesize{\scriptsize}
\tablewidth{0pt}
\tablenum{4}
\tablecaption{The Population of Low-Mass X-ray Binaries in the Galaxy.} 
\tablehead{
 \colhead{Primary} &
 \colhead{Type} &
 \colhead{Number} &
 \colhead{Fraction}}
\startdata
  Neutron Star & Persistent & 46 & 28\% \\
  Neutron Star & Transient & 39 & 23\% \\
  Confirmed BH & Persistent & 0 & 0\% \\
  Confirmed BH & Transient & 16 & 9\% \\
  BH Candidate & Persistent & 2 & 1\% \\
  BH Candidate & Transient & 30 & 18\% \\
  Unidentified & Persistent & 7 & 4\% \\
  Unidentified & Transient & 3 & 2\% \\
  Little Information & Persistent & 17 & 11\% \\
  Little Information & Transient & 7 & 4\%
\enddata
\scriptsize
\label{population}
\end{deluxetable}

%%%%%%%%%%%%%%%%%%%%%%%%%%%%%%%%%%%%%%%%%%%%%%%%%%%%%%%%%%%%%%%%%%%

\acknowledgments

We thank Chris Fryer for stimulating discussions and Ron Remillard for
sharing his private catalogue of black hole candidates. F\"O and DP
thank the ITC at the Harvard-Smithsonian Center for Astrophysics for
their hospitality. F\"O acknowledges support from NSF grant AST
07-08640 and Chandra Theory grant TMO-11003X. DP was supported by the
NSF CAREER award NSF 0746549.


\begin{thebibliography}{99}


\bibitem{LIGO_Virgo} Abadie et al., LIGO Scientific Collaboration, 2010, arXiv:1003.2480 

\bibitem{bailynetal98} Bailyn, C.~D., Jain, R.~K., Coppi, P., \& Orosz, J.~A.\ 1998, 
\apj, 499, 367

\bibitem{bar96} Barret, D., McClintock, J. E., \& Grindlay, J. E. 1996,
  ApJ, 473, 963

\bibitem{bsn97} Beekman, G., et al. 1997, MNRAS, 290, 303

\bibitem{cgb09} Caballero-Nieves et al. 2009, ApJ, 701, 1895

\bibitem{cbm08} Cantrell, A. G., Bailyn, C. D., McClintock, J. E.,
\& Orosz, J. A. 2008, ApJ, 673, L159

\bibitem{cbo10} Cantrell, A. G., et al. 2010, ApJ, 710, 1127 

\bibitem{coz09} Casares, J., et al. 2009, ApJS, 181, 238

\bibitem{chc06} Charles, P. A., \& Coe, M. J. 2006, in Compact Stellar
X-ray Sources, ed. W. Lewin \& M. van der Klis, Cambridge Univ. Press

\bibitem{cow92} Cowley, A. P. 1992, ARAA, 30, 287

\bibitem{cbc10} Crowther, P. A, et al. 2010, MNRAS, 403, L41

\bibitem{dubusetal99} Dubus, G., Lasota, J.-P., Hameury, J.-M., \& 
Charles, P.\ 1999, \mnras, 303, 139 

\bibitem{fen99} Fender, R. P., et al.\ 1999, MNRAS, 304, 865

\bibitem{fen05} Fender, R., Garrington, S., \& Muxlow, T. 2005, ATel,
558

\bibitem{flm99} Filippenko, A. V., et al. 1999, PASP, 111, 969

\bibitem{fcx01} Filippenko, A. V., \& Chornock, R. 2001, IAUC No. 7644

\bibitem{fmh95} Filippenko, A. V., Matheson, T., \& Ho, L. C. 1995, ApJ,
455, 614

\bibitem{fragos10} Fragos, T., Tremmel, M., Rantsiou, E., \& Belczynski, 
K.\ 2010, \apjl, 719, L79

\bibitem{fryer} Fryer, C.~L.\ 1999, \apj, 522, 413

\bibitem{fk01} Fryer, C.~L.\ \& Kalogera, V. 2001, \apj, 554, 548

\bibitem{Fryer02} Fryer, C.~L., Heger, A., Langer, N., \& Wellstein, S.\ 
2002, \apj, 578, 335

\bibitem{geh03} Gelino, D. M., \& Harrison, T. E. 2003, ApJ, 599, 1254

\bibitem{cbk06} Gelino, D. M., et al. 2006, ApJ, 642, 438

\bibitem{gbo01} Greene, J., Bailyn, C. D., \& Orosz, J. A. 2001, ApJ,
554, 1290

\bibitem{gcm01} Greiner, J., Cuby, J. G., \& McCaughrean, M. J. 2001,
Nature, 414, 522

\bibitem{harl04} Harlaftis, E. T., \& Greiner, J. 2004, A\&A, 414, L13

\bibitem{hje95} Hjellming, R. M., \& Rupen, M. P. 1995, Nature, 375, 464

\bibitem{hom06} Homan, J., et al.\ 2006, MNRAS, 366, 235

\bibitem{hyn05} Hynes, R. I. 2005, ApJ, 623, 1026

\bibitem{hynesetal03} Hynes, R.~I., Steeghs, D., Casares, J., Charles, 
P.~A., \& O'Brien, K.\ 2003, \apjl, 583, L95

\bibitem{hyn04} Hynes, R. I., Steeghs, D., Casares, J., Charles, P. A.,
\& O'Brien, K. 2004, ApJ, 609, 317

\bibitem{hughes09} Hughes, S.~A.\ 2009, \araa, 47, 107

\bibitem{jpm09} Johannsen, T., Psaltis, D., \& McClintock, 
J.~E.\ 2009, \apj, 691, 997

\bibitem{kingetal} King, A.~R., Kolb, U., \& Burderi, L.\ 1996, \apjl,
464, L127

\bibitem{kri08} Krimm, H. A., Kennea, J., Barthelmy, S. D., et
al. 2008, ATel, 1610

\bibitem{liu07} Liu, Q. Z., van Paradijs, J. \& van den Heuvel, E. P. J.
2007, A\&A, 469, 807

\bibitem{mar09} Markwardt, C. B., Beardmore, A. P., Miller, J., \&
Swank, J. H. 2009, ATel, 2120

\bibitem{mgc01} McClintock, J. E., et al. 2001, ApJ, 551, L147

\bibitem{mcc06} McClintock, J. E., \& Remillard, R. A. 2006, in Compact
  Stellar X-ray Sources, ed. W. Lewin \& M. van der Klis, CUP

\bibitem{mil09} Miller-Jones, J. C. A., et al. 2009, ApJ, 706, L230

\bibitem{mirrod99} Mirabel, I.~F., \& Rodr{\'{\i}}guez, L.~F.\ 1999, 
\araa, 37, 409

\bibitem{1822} Mu{\~n}oz-Darias, T., Casares, J., \& Mart{\'{\i}}nez-Pais, 
I.~G.\ 2005, \apj, 635, 502

\bibitem{mcm08} --------- 2008, MNRAS, 385, 2205

\bibitem{nm05} Narayan, R., \& McClintock, J.~E.\ 2005, \apj, 623, 1017

\bibitem{narayan91} Narayan, R., Piran, T., \& Shemi, A.\ 1991, 
\apjl, 379, L17

\bibitem{nbc07} Neil, E. T., Bailyn, C. D., \& Cobb, B. E. 2007, apJ,
657, 409

\bibitem{nsv08} Neilsen, J., Steeghs, D., \& Vrtilek, S. D. 2008, MNRAS,
384, 849

\bibitem{oro03} Orosz, J. A. 2003, in A Massive Star Odyssey: From Main
Sequence to Supernova, Proceedings of IAU Symposium No.\ 212, ed. K. van
der Hucht et al., Astron. Soc. Pacific

\bibitem{obm96} Orosz, J. A., Bailyn, C. D., McClintock, J. E., \&
Remillard, R. A. 1996, ApJ, 468, 380

\bibitem{CygX2} Orosz, J.~A., \& Kuulkers, E.\ 1999, \mnras, 305, 132 

\bibitem{oro01} Orosz, J. A., et al.\ 2001, ApJ, 555, 489

\bibitem{oro07} Orosz, J. A., et al. 2007, Nature, 449, 872

\bibitem{oro04} Orosz, J. A., McClintock, J. E., Remillard, R. A., \&
Corbel, S. 2004, ApJ, 616, 376

\bibitem{osm09} Orosz, J. A., et al. 2009, 697, 573

\bibitem{oro10} Orosz, J. A., Steiner, J. F., McClintock, J. E., Torres, M. A. P.,
 Remillard, R. A., \& Bailyn, C. D. 2010, ApJ, submitted

\bibitem{bhevap} Postnov, K.~A., \& Cherepashchuk, A.~M.\ 2003, Astronomy Reports, 47, 989 

\bibitem{pres07} Prestwich A. H., et al. 2007, ApJ, 669, L21

\bibitem{rauetal09} Rau, A., et al.\ 2009, \pasp, 121, 1334

\bibitem{rem06} Remillard, R. A., \& McClintock, J. E. 2006, ARAA, 44,
49

\bibitem{rcf07} Reynolds, M. T., Callanan, P. J., \& Filippenko,
A. V. 2007, MNRAS, 374, 657

\bibitem{smartt09} Smartt, S.\ J.\ 2009, \araa, 47, 63

\bibitem{saf08} Silverman, J. M., \& Filippenko, A. V. 2008, ApJ, 678,
17L

\bibitem{bowen} Steeghs, D., \& Casares, J.\ 2002, \apj, 568, 273 

\bibitem{timmes96} Timmes, F.~X., Woosley, 
S.~E., \& Weaver, T.~A.\ 1996, \apj, 457, 834

\bibitem{tor09} Torres, M. A. P., Jonker, P. G., Steeghs, D., Yan, H.,
Huang, J., \& Soderberg, A. M. 2009, ATel, 2263

\bibitem{vanParadijs} van Paradijs, J.\ 1996, \apjl, 464, L139

\bibitem{wni00} Webb, N. A., et al. 2000, MNRAS, 317, 528

\bibitem{WK99} Wex, N., \& Kopeikin, S.~M.\ 1999, \apj, 514, 388

\bibitem{woosley02} Woosley, S.~E., Heger, 
A., \& Weaver, T.~A.\ 2002, Reviews of Modern Physics, 74, 1015 

\bibitem{zhangetal08} Zhang, W., Woosley, S.~E., \& Heger, A.\ 2008, \apj, 679, 639

\bibitem{zcs03} Zurita, C., Casares, J., \& Shahbaz, T. 2003, ApJ, 582,
369

\bibitem{zsc02} Zurita, C., et al. 2002, MNRAS, 334, 999


\end{thebibliography}
\end{document}